\documentclass[11pt,twoside]{article}

\usepackage{asp2006}
\usepackage{graphicx}

\begin{document}
\title{The continuous star formation history of a giant HII region in M101}   
\author{Rub\'en Garc\'ia-Benito$^{1}$, Enrique P\'erez$^{2}$ and \'Angeles I. D\'iaz$^{1}$}   
\affil{$^{1}$Universidad Aut\'onoma de Madrid, Madrid, Spain}    
\affil{$^{2}$Instituto de Astrof\'\i sica de Andaluc\'\i a (CSIC), Granada, Spain}

\begin{abstract} 
We present results about the star formation process in the giant HII region NGC 5471 in the outskirts of M101. From resolved HST/WPFC2 photometry we find that star formation has been going for the last 70 Ma. We further compare previous results from integrated infrared-optical photometry with the stellar resolved CMD and we discuss the star formation properties of this region and its individual knots, as well as characterizing the different stellar content. This result has very important consequences in our understanding of the burst versus continuous star formation activity in spiral galaxies.
\end{abstract}



\section{Images and reduction}   
The WFPC2 data was retrieved from the HST archive. The images were taken through two emission-line filters, F656N 
(H$\alpha$) and F673N ([SII]), and two continuum filters, F547M and F675W. The pipeline 
processed HST images were subjected to the usual processing using the IRAF and STSDAS software. We have 
subtracted  scaled H$\alpha$ and [SII] images from the continuum F675W image obtining a new  almost emission line 
free image (F675W’).

We obteined JHK$_{S}$ CCD images with the TNG 3.58 m at la Palma with the ARNICA camera on April 28 (2002), with 
a seeing of 0.8 arcsec. 

\section{Integrated Study}
In order to perform an analysis of the stellar population content of the main components in NGC 5471 we derived the 
integrated magnitudes (JHK + F547M + F675W’) and colors in eleven different apertures by means of the \texttt{polyphot} 
task in IRAF. They were chosen and defined in the H contour plot, maximizing the area in order to include the features 
of all images.


To find the correspondence between the observed photometric properties and the models, we have used CHORIZOS, a 
code written in IDL by Maíz-Apellániz \citet{maiz}. The code uses $\chi^{2}$ minimization to find all models (Starburst99) 
compatible with the observed data in the 4-dimensional parameter space. We have fixed two of these parameters: the 
known metallicity of NGC 5471 and the type of dust R$_{V}$ = 3.1, leaving unconstrained the amount of extinction and 
the age.

The most probable extinction in each knot is plotted against its age in figure \ref{cmd}. 
For some knots more than one solution is found. The general behaviour is as expected: an old knot shows a 
small amount of extinction. 

\section{Stellar Study}
The stellar photometric analysis was performed with the HSTphot package \citet{dolphin}. 
 
We obtained the completeness limit by adding artificial stars to the images and subject them to the same analysis as 
the real ones. The 50\% completeness of the F547M filter derived on the basis of the CMD data is 25.2 mag, while 
that for the F657W' filter is 50\% at 24.7 mag.

\begin{figure}
\centering
\includegraphics[angle=-90, width=6.5cm]{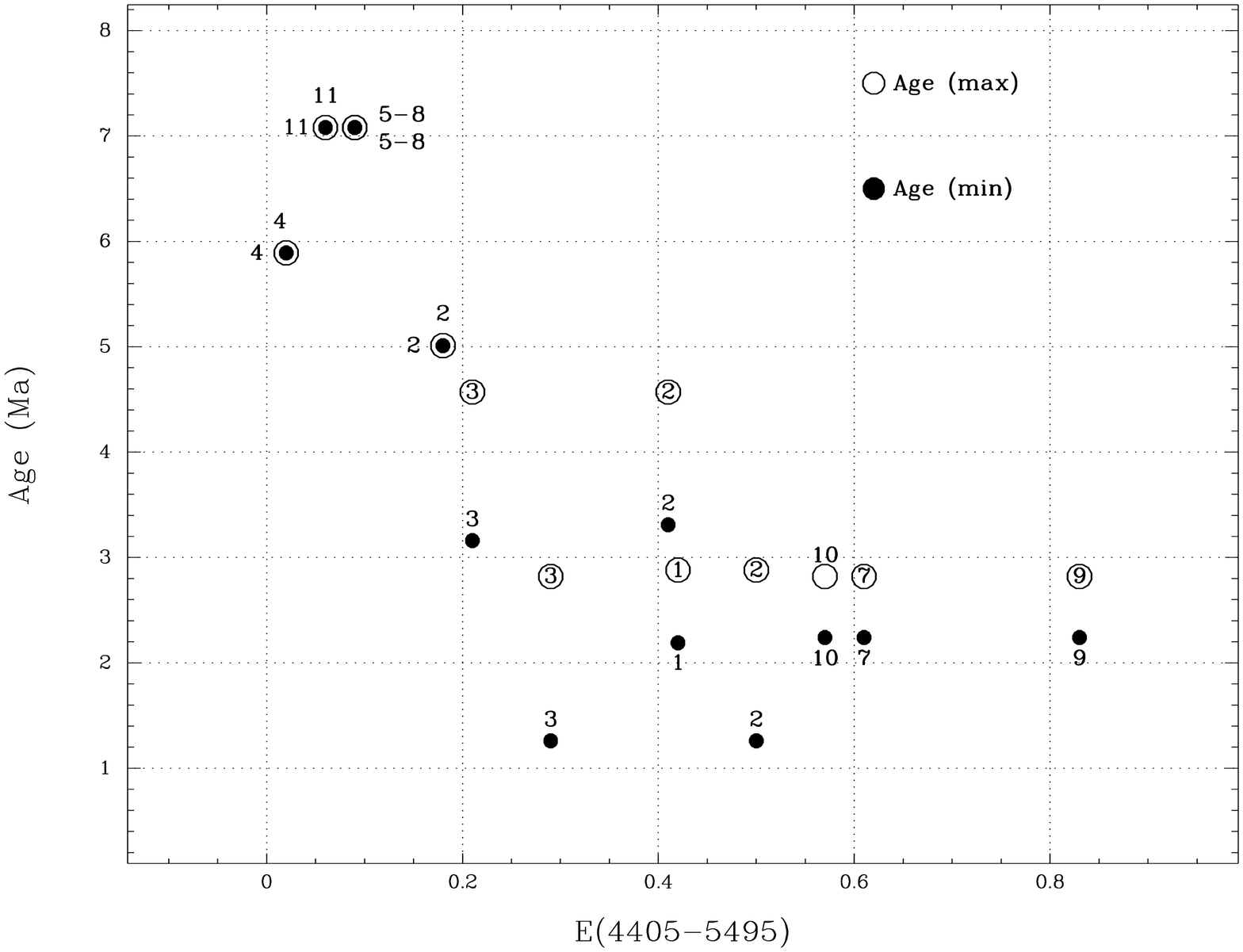}
\includegraphics[angle=-90, width=6.5cm]{ruben_fig2.eps}
\caption{On the left, most probable extintion in each knot against its age. On the right, CMD of NGC5471 with 
isochrones from 4 to 70 Ma.}
\label{cmd}
\end{figure}

Our CMD consist of the sum of the intrinsic population IP and the contaminating field population FS, and therefore the 
CMD shows both populations (IP+FS). The field star contribution has been estimated using the CMD of adjacent fields 
FS' cotaining only foreground and background stars. FS contribution may be removed statistically \citep{bella} by 
comparing the local density of stars in the two diagrams. 

We assume a distance modulus of (m−M) = 29.3 for M101 \citep{stetson}. In Figure \ref{cmd} is plotted the final 
CMD with isochrones provided by Miguel Cervi\~no, showing a mixture of different age populations ranged from 4 to 
70 Ma: the star formation in NGC 5471 has been going for at least that period.



\end{document}